\documentclass[12pt]{article}
\usepackage{hyperref}
\usepackage[english]{babel}
\usepackage{amsmath}
\usepackage{latexsym}
\usepackage{amssymb}
\usepackage[all]{xy}
\usepackage[dvips]{graphicx}
\usepackage{epsfig}
\usepackage{psfrag}

\numberwithin{equation}{section} \numberwithin{table}{section}
\numberwithin{figure}{section}

\begin{document}



\begin{titlepage}
   \begin{flushright}
{\small MPP-2013-277}
  \end{flushright}

   \begin{center}

     \vspace{20mm}

     {\LARGE \bf Entanglement thermodynamics for non-conformal D-branes

     \vspace{3mm}}

     \vspace{10mm}

    Da-Wei Pang

     \vspace{5mm}

      {\small \sl Max-Planck-Institut f\"{u}r Physik (Werner-Heisenberg-Institut)\\
      F\"{o}hringer Ring 6, 80805 M\"{u}nchen, Germany}\\

     {\small \tt dwpang@mppmu.mpg.de}
     \vspace{10mm}

   \end{center}

\begin{abstract}
\baselineskip=18pt
We study thermodynamics of entanglement entropy for weakly excited states in certain non-conformal fields theories, whose gravity duals are given by non-conformal Dp-branes. We observe that the entanglement entropy of a sufficiently small system in non-conformal backgrounds still obeys a first-law like relation, just as the AdS counterparts investigated in arXiv: 1212.1164 [hep-th]. The effective temperature is proportional to the inverse of the size of the subsystem. The proportionality is a dimensionless constant which is only determined by the shape of the entangling region and independent of any coupling. This universality is confirmed by working with the ten-dimensional string frame metric as well as the lower-dimensional effective metric. When the entangling region is a strip and translational invariance is broken by metric fluctuations, we derive a first-law like relation where additional components of the stress energy tensor are involved.
\end{abstract}
\setcounter{page}{0}
\end{titlepage}

\pagestyle{plain} \baselineskip=19pt

\tableofcontents

\section{Introduction}
The AdS/CFT correspondence~\cite{Maldacena:1997re, Aharony:1999ti} establishes fundamental equivalence between weakly coupled gravity theories and strongly coupled quantum field theories, hence it provides powerful tools for investigating the dynamics of strongly coupled systems in the real world. One remarkable example of the success of AdS/CFT is the holographic calculation of the entanglement entropy~\cite{Ryu:2006bv, Ryu:2006ef}, which states that when the bulk theory is Einstein gravity, the holographic entanglement entropy (HEE) is given by
\begin{equation}
S_{A}=\frac{{\rm Area}(\gamma)}{4G_{N}},
\end{equation}
where $\gamma$ denotes the minimal surface and $G_{N}$ is the Newton constant. This relation has been proven in~\cite{Casini:2011kv} for a spherical entangling region and in~\cite{Lewkowycz:2013nqa} for more general cases. HEE at finite temperature has also been extensively studied, e.g. in~\cite{Faraggi:2007fu, Bah:2008cj, Fischler:2012ca, Fischler:2012uv}. For reviews on this fascinating subject, see~\cite{Nishioka:2009un, Takayanagi:2012kg}.

The entanglement entropy is a good measure of quantum information for a pure state and plays an important role in quantum many-body physics. Generically, the amount of information included in a system can be related to the total energy of that system by the first law of thermodynamics, $dE=TdS$. Therefore it is natural to ask if there exists an analogous relation for the entanglement entropy. It was first observed in~\cite{Bhattacharya:2012mi} that when the size of the subsystem is small, a first-law like relation between the entanglement entropy and the total energy for excited states can be derived
\begin{equation}
\Delta E=T_{\rm ent}\Delta S_{A},
\end{equation}
where $T_{\rm ent}$ is the entanglement temperature. The original investigations focused on static and translationally invariant examples and a further proof of this first-law like relation with time-dependent excitations for spherical subsystems was provided in~\cite{Nozaki:2013vta}. Subsequently it was found in~\cite{Guo:2013aca, Allahbakhshi:2013rda, Blanco:2013joa} that when the subsystem is a strip, the first-law like relation involves additional components of the energy momentum tensor other than $T_{tt}$. For related references on this topic, see~\cite{Wong:2013gua, Caputa:2013eka, He:2013rsa, Caputa:2013lfa}.

The derivations on the first-law like relation of HEE~\cite{Bhattacharya:2012mi, Nozaki:2013vta} were performed in asymptotically AdS spacetimes, that is, the corresponding ground states are described by CFTs. However, no matter the system is conformal or not, the first law of thermodynamics always holds. Therefore one may ask if we can obtain a similar relation for the HEE in non-conformal systems. In particular, the entanglement temperature scales as
\begin{equation}
T_{\rm ent}=\frac{c}{l},
\end{equation}
where $c$ is a dimensionless constant and $l$ denotes the size of the subsystem. This behavior can be fixed by dimensional analysis. The authors of~\cite{He:2013rsa} considered HEE of excited states in Einstein-Maxwell-dilaton theory and found that $c$ is given by the ratio of two dimensionful constants with the same dimension. It would be interesting to see if this is a universal behavior for non-conformal systems.

We consider the entanglement thermodynamics of a particular class of non-conformal systems, that is, non-conformal D-branes. The connections between non-conformal D-branes and their field theory duals were successfully established in~\cite{Itzhaki:1998dd}. The validity of the gravity description is controlled by a dimensionless effective coupling $g_{\rm eff}$ and it is natural to speculate that here $c$ may be proportional to $g_{\rm eff}$ or the ratio of two dimensionful couplings with the same dimension. However, we find that for non-conformal D-branes, $c$ is still independent of any gauge coupling. This observation is confirmed by performing calculations using both ten-dimensional string frame metric and the lower dimensional effective metric, which also exhibits its universality. Moreover, when the entangling region is a strip, we derive a first-law like relation for the HEE of excited states by taking advantage of the holographic renormalization of non-conformal D-branes~\cite{Kanitscheider:2008kd}.

The rest of the paper is organized as follows: We review the derivation of the first-law like relation in asymptotically AdS spacetimes and some backgrounds of non-conformal D-branes in Section 2. Then in Section 3 we calculate the entanglement temperature using ten-dimensional string frame metric. In Section 4 we revisit the entanglement temperature by working with the lower dimensional effective metric and confirm its universality. A first-law like relation for the HEE of excited states is obtained in Section 5, where other components of the energy-momentum tensor other than $T_{tt}$ are involved. A summary and discussion will be given in Section 6.
\section{Preliminaries}
In this section we review some backgrounds on entanglement thermodynamics and non-conformal D-branes. In the first subsection we present a brief overview on the computation of entanglement temperature and the derivation of a first-law like relation in asymptotically AdS spacetimes~\cite{Bhattacharya:2012mi, Nozaki:2013vta}. In the second subsection we introduce the essential physics of non-conformal D-branes.
\subsection{The first-law like relation of HEE in AdS}
\label{1stAdS}
First let us consider asymptotically $AdS_{d+1}$ background,
\begin{equation}
ds^{2}=\frac{R^{2}}{z^{2}}\left[-f(z)dt^{2}+g(z)dz^{2}+\sum\limits^{d-1}_{i=1}dx_{i}^{2}\right].
\end{equation}
The asymptotic boundary is located at $z\rightarrow0$, where we can approximate $g(z)=1/f(z)\approx1+mz^{d}$. $m$ is a constant which is related to the holographic stress tensor by~\cite{Balasubramanian:1999re}
\begin{equation}
T_{tt}=\frac{(d-1)}{16\pi G_{N}}R^{d-1}m.
\end{equation}
An important assumption in~\cite{Bhattacharya:2012mi} is that the size of the subsystem is so small that $ml^{d}\ll1$, which means the minimal surface is localized near the asymptotic boundary. Therefore we do not care about details of the IR region $z\rightarrow\infty$, for example, we can have black branes which correspond to thermal states or stars which correspond to zero temperature states in the field theory side.

The holographic entanglement entropy (HEE) in Einstein gravity for a subsystem $A$ is given by~\cite{Ryu:2006bv, Ryu:2006ef}
\begin{equation}
S_{A}=\frac{{\rm Area(\gamma)}}{4G_{N}},
\end{equation}
where $\gamma$ denotes the minimal area surface whose boundary coincides with $\partial A$. Since $\gamma$ satisfies the extremal surface condition in pure AdS, the first order correction to HEE under a metric perturbation can be conveniently evaluated as follows~\cite{Nozaki:2013vta}:
\begin{itemize}
\item Starting with the extremal surface $\gamma$ of pure AdS and calculating its area;
\item Evaluating the area of the same $\gamma$ in the perturbed metric;
\item Finally taking the difference between the two areas.
\end{itemize}
Therefore the first order correction to HEE can be expressed as
\begin{equation}
\label{1sthee}
\Delta S_{A}=\frac{1}{8G_{N}}\int d^{d-1}\xi\sqrt{{\rm det}G^{(0)}}G^{(1)}_{\alpha\beta}G^{(0)\alpha\beta},
\end{equation}
where $\xi$ denotes the coordinates on the $(d-1)$-dimensional extremal surface and $G^{(0)}, G^{(1)}$ are the induced metric on the minimal surface with respect to the ground state metric and its first order perturbations. Given these formulae the first-law like relation for the HEE reads
\begin{equation}
\Delta E_{A}=T_{\rm ent}\Delta S_{A},
\end{equation}
where $T_{\rm ent}=c/l$ with $l$ being the size of $A$ and
\begin{eqnarray}
\label{cAdS}
c&=&\frac{2(d^{2}-1)\Gamma(\frac{1}{2}+\frac{1}{d-1})\Gamma(\frac{d}{2(d-1)})^{2}}{\sqrt{\pi}\Gamma(\frac{1}{d-1})\Gamma({\frac{1}{2(d-1)}})^{2}},~~~{\rm strip}\nonumber\\
c&=&\frac{d+1}{2\pi},~~~{\rm sphere}.
\end{eqnarray}
$\Delta E_{A}$ denotes the total energy in $A$, which is given by
\begin{equation}
\Delta E_{A}=\int_{A} T_{tt},
\end{equation}
where $T_{tt}$ is the $tt$-component of the energy stress tensor. It can be seen from~(\ref{cAdS}) that $c$ is universal when the shape of the the subsystem $A$ is fixed.

Moreover, for a more general class of asymptotically AdS spacetimes written in the Fefferman-Graham (FG) gauge,
\begin{equation}
ds^{2}=\frac{R^{2}}{z^{2}}(dz^{2}+g_{\mu\nu}(z,x)dx^{\mu}dx^{\nu}),
\end{equation}
the first-law like relation receives contributions from other components of the energy-momentum tensor besides $T_{tt}$~\cite{Guo:2013aca, Allahbakhshi:2013rda, Blanco:2013joa},
\begin{equation}
\Delta E_{A}=T_{\rm ent}\Delta S_{A}+\frac{d-1}{d+1}\Delta PV_{d-1},
\end{equation}
where $\Delta P=T_{xx}$ and $V_{d-1}$ denotes the volume of $A$.
\subsection{Non-conformal D-branes}
The main objective of this paper is to investigate the entanglement thermodynamics in non-conformal backgrounds, hence a class of natural candidates are non-conformal Dp-branes. In this subsection we will review some basic properties of non-conformal Dp-branes for later convenience. Consider $N$ coincident extremal Dp-branes in string frame~\cite{Itzhaki:1998dd}
\begin{eqnarray}
& &ds^{2}=H_{p}^{-1/2}(-dt^{2}+\sum\limits^{p}_{i=1}dx_{i}^{2})+H_{p}^{1/2}(dr^{2}+r^{2}d\Omega^{2}_{8-p}),\nonumber\\
& &H_{p}=1+\frac{d_{p}g_{YM}^{2}N}{\alpha^{\prime2}U^{7-p}},~~U=\frac{r}{\alpha^{\prime}},~~d_{p}=2^{7-2p}\pi^{\frac{9-3p}{2}}\Gamma(\frac{7-p}{2}),\nonumber\\
& &e^{\phi}=H_{p}^{(3-p)/4},
\end{eqnarray}
where $g_{YM}$ denotes the Yang-Mills coupling. After taking the field theory limit
\begin{equation}
g_{YM}^{2}=(2\pi)^{p-2}g_{s}\alpha^{\prime (p-3)/2}={\rm fixed},~~U~{\rm fixed},~~\alpha^{\prime}\rightarrow0,
\end{equation}
the string frame metric becomes
\begin{eqnarray}
\label{Dp}
& &ds^{2}=\alpha^{\prime}\left(\frac{U^{(7-p)/2}}{g_{YM}\sqrt{d_{p}N}}(-dt^{2}+\sum\limits^{p}_{i=1}dx_{i}^{2})+\frac{g_{YM}\sqrt{d_{p}N}}{U^{(7-p)/2}}dU^{2}
+g_{YM}\sqrt{d_{p}N}U^{(p-3)/2}d\Omega_{8-p}^{2}\right),\nonumber\\
& &e^{\phi}=(2\pi)^{2-p}g_{YM}^{2}\left(\frac{g_{YM}^{2}d_{p}N}{U^{7-p}}\right)^{\frac{3-p}{4}}.
\end{eqnarray}
In the following we will work in the unit $\alpha^{\prime}=1$. Note that the supergravity description is valid when both the curvature and the dilaton are small, which can be expressed in terms of the effective dimensionless coupling constant $g_{\rm eff}$ as
\begin{equation}
1\ll g_{\rm eff}\ll N^{\frac{4}{7-p}},~~g_{\rm eff}^{2}\approx g_{YM}^{2}NU^{p-3}.
\end{equation}
Since we are interested in non-conformal Dp-branes with $p=1,2,4$, the supergravity description can be trusted when
\begin{eqnarray}
\label{bound}
& &g_{YM}N^{1/6}\ll U\ll g_{YM}\sqrt{N},~~{\rm for~ D1 ~branes},\nonumber\\
& &g_{YM}N^{1/5}\ll U\ll g_{YM}^{2}N,~~{\rm for~D2~branes},\nonumber\\
& &N^{-1}\ll g_{YM}^{2}U\ll N^{1/3},~~{\rm for~D4 ~branes}.
\end{eqnarray}
Moreover, the near-extremal Dp-brane solutions in the same limit can be expressed as
\begin{eqnarray}
\label{NextDp}
ds^{2}&=&\alpha^{\prime}[\frac{U^{(7-p)/2}}{g_{YM}\sqrt{d_{p}N}}(-f(U)dt^{2}+\sum\limits^{p}_{i=1}dx_{i}^{2})
+\frac{g_{YM}\sqrt{d_{p}N}}{f(U)U^{(7-p)/2}}dU^{2}\nonumber\\
& &+g_{YM}\sqrt{d_{p}N}U^{(p-3)/2}d\Omega_{8-p}^{2}],\nonumber\\
f(U)&=&1-\frac{U_{0}^{7-p}}{U^{7-p}},
\end{eqnarray}
while the dilaton remains the same as in the extremal counterparts.
\section{The entanglement temperature in string frame}
In this section we calculate the entanglement temperature using string frame metric. Our starting point is a slight modification of the near-extremal Dp-brane metric~(\ref{NextDp}),
\begin{eqnarray}
\label{mstr}
ds^{2}&=&\alpha^{\prime}[\frac{U^{(7-p)/2}}{g_{YM}\sqrt{d_{p}N}}(-f(U)dt^{2}+\sum\limits^{p}_{i=1}dx_{i}^{2})
+\frac{g(U)g_{YM}\sqrt{d_{p}N}}{U^{(7-p)/2}}dU^{2}\nonumber\\
& &+g_{YM}\sqrt{d_{p}N}U^{(p-3)/2}d\Omega_{8-p}^{2}],\nonumber\\
g(U)&=&1/f(U)\approx1+\frac{U_{0}^{7-p}}{U^{7-p}},
\end{eqnarray}
which is taken as the metric describing the excited states in a broader sense, following~\cite{Bhattacharya:2012mi}. In particular, we consider the metric to describe either a black brane or a star, which corresponds to thermal state or zero temperature state respectively in the field theory side. Moreover, we assume that the size of the subsystem is sufficiently small, which will enable us to carry out the analysis in a parallel way to the AdS cases~\cite{Bhattacharya:2012mi}.

Consider the case where the entangling region is a strip, $$x_{1}\in [-l/2,l/2],~~x_{i}\in[0, L],~i=2,\cdots,p.$$ The induced metric is given by
\begin{eqnarray}
ds^{2}_{\rm ind}&=&\frac{g_{YM}\sqrt{d_{p}N}}{U^{(7-p)/2}}\left[g(U)+\frac{U^{7-p}}{g_{YM}^{2}d_{p}N}x^{\prime2}(U)\right]dU^{2}\nonumber\\
& &+\frac{U^{(7-p)/2}}{g_{YM}\sqrt{d_{p}N}}\sum\limits^{p}_{i=2}dx_{i}^{2}+g_{YM}\sqrt{d_{p}N}U^{(p-3)/2}d\Omega_{8-p}^{2},
\end{eqnarray}
where $x(U)\equiv x_{1}(U)$ and the prime denotes derivative with respect to $U$. Recall that the metric~(\ref{mstr}) is written in string frame, hence the expression for HEE should be modified as~\cite{Ryu:2006ef}
\begin{equation}
S_{A}=\frac{1}{4G_{N}^{(10)}}\int d^{8}x\sqrt{{\rm det}g_{\rm ind}}e^{-2\phi},
\end{equation}
where $\phi$ is the dilaton.

It should be pointed out that the HEE of non-conformal Dp-branes with a strip entangling region was investigated in~\cite{vanNiekerk:2011yi} and here we just rewrite the results for completeness. The boundary separation length $l$ is
\begin{eqnarray}
l&=&2g_{YM}\sqrt{d_{p}N}\int dU\frac{U_{\ast}^{(9-p)/2}}{U^{8-p}}\frac{1}{\sqrt{1-\frac{U_{\ast}^{9-p}}{U^{9-p}}}}\nonumber\\
&=&\frac{4\sqrt{\pi}}{5-p}\frac{\Gamma(\frac{7-p}{9-p})}{\Gamma(\frac{5-p}{2(9-p)})}\frac{g_{YM}\sqrt{d_{p}N}}{U_{\ast}^{(5-p)/2}},
\end{eqnarray}
where $U_{\ast}$ denotes the turning point of the extremal surface where $x^{\prime}$ diverges. The HEE is given by
\begin{equation}
S_{A}=\frac{1}{4G_{N}^{(10)}}(2\pi)^{2p-4}L^{p-1}\Omega_{8-p}g_{YM}^{-2}d_{p}N(U_{\rm max}^{2}-\frac{\sqrt{\pi}\Gamma(\frac{7-p}{9-p})}{\Gamma(\frac{5-p}{2(9-p)})}U_{\ast}^{2}),
\end{equation}
where $U_{\rm max}$ is the UV cutoff. The finite part (the second term in the above expression) can be rewritten in terms of $l$,
\begin{eqnarray}
S_{A{\rm finite}}&=&-\frac{1}{4G_{N}^{(10)}}\Omega_{8-p}L^{p-1}\lambda_{Dp}g_{YM}^{\frac{2(p-3)}{5-p}}N^{\frac{7-p}{5-p}}l^{\frac{4}{p-5}},\nonumber\\
\lambda_{Dp}&=&(2\pi)^{2p-4}\pi^{\frac{9-p}{2(5-p)}}(\frac{4}{5-p})^{4/(5-p)}(\frac{\Gamma(\frac{7-p}{9-p})}{\Gamma(\frac{5-p}{2(9-p)})})^{\frac{9-p}{5-p}}
d_{p}^{\frac{7-p}{5-p}},
\end{eqnarray}
Before proceeding, it should be emphasized that during the calculations we keep $U_{\ast}$ much larger than the lower bound in~(\ref{bound}) and $U_{\rm max}$ much smaller than the the upper bound, so that the supergravity description is always valid.

Next we can compute the first order correction to HEE following the steps introduced in Section~\ref{1stAdS},
\begin{eqnarray}
\Delta S_{A}&=&\frac{1}{8G_{N}^{(10)}}\int d^{8}x\sqrt{G^{(0)}}G^{(1)}_{\alpha\beta}G^{(0)\alpha\beta}e^{-2\phi}\nonumber\\
&=&\frac{1}{4G_{N}^{(10)}}\Omega_{8-p}L^{p-1}g_{YM}^{-4}U_{0}^{7-p}c_{S_{A}}l^{2},\nonumber\\
c_{S_{A}}&=&\frac{(5-p)^{2}}{16(19-3p)\sqrt{\pi}}\frac{\Gamma(\frac{5-p}{9-p})}{\Gamma(\frac{5-p}{9-p}+\frac{1}{2})}
\frac{\Gamma(\frac{5-p}{2(9-p)})^{2}}{\Gamma(\frac{7-p}{9-p})^{2}}.
\end{eqnarray}
On the other hand, the energy of the excited states can be evaluated by making use of the D-brane energy obtained in~\cite{Myers:1999psa}, which gives
\begin{equation}
\Delta E=\frac{1}{4G_{N}^{(10)}}\frac{9-p}{2}\Omega_{8-p}g_{YM}^{-4}U_{0}^{7-p}L^{p-1}l.
\end{equation}
Finally the entanglement temperature is given by
\begin{equation}
T_{\rm ent}=\frac{\Delta E}{\Delta S_{A}}=\frac{c_{Dp}}{l},~~c_{Dp}=\frac{9-p}{8c_{S_{A}}}.
\end{equation}
It can be seen that the entanglement temperature is still given by $c/l$, where $c$ is a constant independent of any coupling. In other words, for non-conformal Dp-branes the constant $c$ is still universal once the shape of the subsystem is fixed, in contrary to the cases studied in~\cite{He:2013rsa}.
\section{Universality of the entanglement temperature}
In the previous section we have seen that when the entangling region is a strip, the entanglement temperature calculated in string frame exhibits certain universality, that is, it does not depend on any coupling. In order to see if the entanglement temperature possesses the universality for more general situations, we calculate the entanglement temperature in $(p+2)$-dimensional backgrounds, which can be obtained by standard Kaluza-Klein reduction. We consider the cases with the entangling region being a strip and a sphere and it turns out that the entanglement temperature is still independent of the couplings.

The $(p+2)$-dimensional effective description of non-conformal Dp-branes is given by Einstein-dilaton theory~\cite{Boonstra:1998mp},
\begin{eqnarray}
S&=&\frac{N^{2}}{16\pi G_{N}^{(p+2)}}\int d^{p+2}x\sqrt{-g}[R-\frac{1}{2}(\partial\Phi)^{2}+V(\Phi)]\nonumber\\
& &-\frac{N^{2}}{8\pi G_{N}^{(p+2)}}\int d^{p+1}x\sqrt{-h}K,\nonumber\\
V(\Phi)&=&\frac{1}{2}(9-p)(7-p)N^{-2\lambda/p}e^{a\Phi},\nonumber\\
\Phi&=&\frac{2\sqrt{2(9-p)}}{\sqrt{p}(7-p)}\phi,~~a=-\frac{\sqrt{2}(p-3)}{\sqrt{p(9-p)}},\nonumber\\
\lambda&=&2(p-3)/(7-p).
\end{eqnarray}
The corresponding black brane solution reads
\begin{eqnarray}
\label{p+2}
ds^{2}_{p+2}&=&(Ne^{\phi})^{2\lambda/p}\left[u^{2}(-f(u)dt^{2}+\sum\limits_{i=1}^{p}dx_{i}^{2})+\frac{du^{2}}{u^{2}f(u)}\right],\nonumber\\
f(u)&=&1-\left(\frac{u_{0}}{u}\right)^{2(7-p)/(5-p)},~~e^{\phi}=\frac{1}{N}(g_{YM}^{2}N)^{\frac{7-p}{2(5-p)}}u^{\frac{(p-7)(p-3)}{2(p-5)}},
\end{eqnarray}
where $u_{0}$ is related to $U_{0}$ in~(\ref{NextDp}) by $u_{0}^{2}=(g_{YM}^{2}N)^{-1}U_{0}^{5-p}$. It can be seen that the metric~(\ref{p+2}) is conformal to $AdS_{p+2}$ with a $u$-dependent warp factor. Note that our main objective is to check the universality of the entanglement temperature, hence we have already
neglected the numerical factors in the relevant configurations. To be explicit we study the D1-, D2- and D4-branes separately.
\subsection{The strip}
First let us consider the case where the entangling region is a strip.  Here we still assume that~(\ref{p+2}) describes the gravity dual of the excited states in a broader sense, that is, it can represent both black branes and stars. Moreover, since the dilaton remains the same in both the zero temperature background and the black brane background, it is sufficient to consider the metric fluctuations when evaluating the entanglement entropy.

\textbf{D1-branes}

We rewrite the metric~(\ref{p+2}) for D1-branes as follows for convenience,
\begin{equation}
ds^{2}_{D1}=\frac{1}{g_{YM}^{2}Nz^{2}}\left[\frac{1}{z^{2}}(-f(z)dt^{2}+dx^{2})+\frac{dz^{2}}{z^{2}f(z)}\right],~~~f(z)=1-\frac{z^{3}}{z_{0}^{3}},
\end{equation}
where we have introduced $u=1/z,~u_{0}=1/z_{0}$. The HEE is given by
\begin{equation}
S_{D1}=\frac{1}{2G_{N}^{(3)}}\frac{N^{3/2}}{g_{YM}}\int dz\frac{1}{z^{2}}\sqrt{g(z)+x^{\prime2}(z)}.
\end{equation}
Note that the integral is exactly the same as that for $AdS_{4}$. The boundary separation length can be obtained in a similar way,
\begin{equation}
\label{ld1}
l=\frac{2\sqrt{\pi}\Gamma(\frac{3}{4})}{\Gamma(\frac{1}{4})}z_{\ast},
\end{equation}
where $z_{\ast}$ still denotes the turning point.
For completeness we caculate the leading order HEE
\begin{equation}
S^{(0)}_{D1}=\frac{N^{3/2}}{2G_{N}^{(3)}g_{YM}}\left(\frac{1}{\epsilon}-2\pi\frac{\Gamma(\frac{3}{4})^{2}}{\Gamma(\frac{1}{4})^{2}}\frac{1}{l}\right),
\end{equation}
which agrees with the result in previous section up to some numerical factors.
The first order correction to HEE can be evaluated by using~(\ref{1sthee}),
\begin{equation}
\Delta S_{D1}=\frac{N^{3/2}}{128G_{N}^{3}g_{YM}}\frac{\Gamma(\frac{1}{4})^{2}}{\Gamma(\frac{3}{4})^{2}}\frac{l^{2}}{z_{0}^{3}}.
\end{equation}
On the other hand, even though the background~(\ref{p+2}) is not asymptotically AdS, it is still possible to construct the stress-energy tensor ~\cite{Cai:1999xg}, which enables us to read off the energy of the excited states,
\begin{equation}
\Delta E=\frac{1}{4\pi G_{N}^{(3)}}g_{YM}^{-4}U_{0}^{6}l.
\end{equation}
Finally we obtain the entanglement temperature
\begin{equation}
\label{D1ld}
T_{\rm ent}=\frac{\Delta E}{\Delta S_{D1}}=\frac{c_{D1}}{l},~~c_{D1}=\frac{32}{\pi}\frac{\Gamma(\frac{3}{4})^{2}}{\Gamma(\frac{1}{4})^{2}}.
\end{equation}
\textbf{D2-branes}

The effective D2-brane metric is given by
\begin{eqnarray}
ds^{2}_{D2}&=&(g_{YM}^{2}N)^{-1/3}z^{-1/3}\left[\frac{1}{z^{2}}(-f(z)dt^{2}+\sum\limits_{i=1}^{2}dx_{i}^{2})+\frac{dz^{2}}{z^{2}f(z)}\right],\nonumber\\
f(z)&=&1-\left(\frac{z}{z_{0}}\right)^{10/3}.
\end{eqnarray}
We can easily work out the minimal surface
\begin{equation}
A=2Lg_{YM}^{-2/3}N^{5/3}\int dz\frac{\sqrt{g(z)+x^{\prime2}}}{z^{7/3}},~~~g(z)=\frac{1}{f(z)}=1+\left(\frac{z}{z_{0}}\right)^{10/3}.
\end{equation}
Note that the integral is no longer the same as the corresponding $AdS$ counterpart.
The boundary separation length can be obtained in a similar way,
\begin{equation}
\label{ld2}
l=\frac{3\sqrt{\pi}}{7}\frac{\Gamma(\frac{5}{7})}{\Gamma(\frac{17}{14})}z_{\ast}.
\end{equation}
We can also evaluate the the leading order HEE as well as the leading order correction,
\begin{eqnarray}
S_{D2}^{(0)}&=&\frac{3}{8G_{N}^{(4)}}g_{YM}^{-\frac{2}{3}}N^{\frac{5}{3}}L\left(\frac{1}{\epsilon^{4/3}}-\frac{c_{D2}^{(0)}}{l^{4/3}}\right),\nonumber\\
c_{D2}^{(0)}&=&\pi^{\frac{7}{6}}\left(\frac{3}{7}\right)^{\frac{4}{3}}\frac{\Gamma(\frac{5}{7})^{\frac{7}{3}}}
{\Gamma(\frac{3}{14})\Gamma(\frac{17}{14})^{\frac{4}{3}}},
\end{eqnarray}
\begin{eqnarray}
\Delta S_{D2}&=&\frac{c^{(1)}_{D2}}{4G_{N}^{(4)}}g_{YM}^{-\frac{2}{3}}N^{\frac{5}{3}}L\frac{l^{2}}{z_{0}^{\frac{10}{3}}},\nonumber\\
c^{(1)}_{D2}&=&\frac{49}{9\sqrt{\pi}}\frac{\Gamma(\frac{3}{7})\Gamma(\frac{17}{14})^{2}}{\Gamma(\frac{27}{14})\Gamma(\frac{5}{7})^{2}}.
\end{eqnarray}
The energy of excited states can be obtained using the method in~\cite{Cai:1999xg},
\begin{equation}
\Delta E=\frac{1}{16\pi G_{N}^{(4)}}\frac{7}{2}g_{YM}^{-4}U_{0}^{5}Ll.
\end{equation}
Finally we obtain the entanglement temperature
\begin{equation}
T_{\rm ent}=\frac{\Delta E}{\Delta S_{D2}}=\frac{c_{D2}}{l},~~c_{D2}=\frac{7}{8\pi c^{(1)}_{D2}},
\end{equation}
which shows that the entanglement temperature does not depend on any coupling.

\textbf{D4-branes}

The calculations for D4-branes can be performed in a parallel way, so we will be in brief.
Given the metric
\begin{eqnarray}
ds^{2}&=&\frac{g_{YM}\sqrt{N}}{z^{1/2}}\left[\frac{1}{z^{2}}(-f(z)dt^{2}+\sum\limits_{i=1}^{4}dx_{i}^{2})+\frac{dz^{2}}{z^{2}f(z)}\right],\nonumber\\
f(z)&=&1-\left(\frac{z}{z_{0}}\right)^{6},
\end{eqnarray}
it is straightforward to obtain the minimal surface area
\begin{equation}
A=2g_{YM}^{2}N^{3}L^{3}\int dz\frac{\sqrt{g(z)+x^{\prime2}}}{z^{5}}.
\end{equation}
Note that here the integral is exactly the same as the $AdS_{7}$ case. The boundary separation length is given by
\begin{equation}
\label{ld4}
l=\frac{2\sqrt{\pi}\Gamma(\frac{3}{5})}{\Gamma(\frac{1}{10})}z_{\ast}.
\end{equation}
The leading order correction to HEE is
\begin{eqnarray}
\Delta S_{D4}&=&\frac{c^{(1)}_{D4}}{224G_{N}^{(6)}}g_{YM}^{2}N^{3}L^{3}l^{2},\nonumber\\
c^{(1)}_{D4}&=&\frac{\Gamma(\frac{1}{5})\Gamma(\frac{1}{10})^{2}}{\sqrt{\pi}\Gamma(\frac{7}{10})\Gamma(\frac{3}{5})^{2}}.
\end{eqnarray}
The energy of excited states is given by
\begin{equation}
\Delta E=\frac{1}{16\pi G_{N}^{6}}\frac{5}{2}g_{YM}^{-4}U_{0}^{3}L^{3}l.
\end{equation}
Therefore the entanglement temperature can be read off as
\begin{equation}
T_{\rm ent}=\frac{\Delta E}{\Delta S_{D4}}=\frac{c_{D4}}{l},~~c_{D4}=\frac{35}{\pi c^{(1)}_{D4}},
\end{equation}

\textbf{Remarks}

Here are some remarks on the results we have obtained in this subsection:
\begin{itemize}
\item The entanglement temperature in all the examples we study scales as $c/l$, where $c$ is independent of any coupling. In other words, the constant $c$ is universal once the shape of the subsystem is fixed.
\item This is the same as the AdS cases studied in~\cite{Bhattacharya:2012mi}, even though here the backgrounds do not possess conformal symmetry.
\item For D1- and D4- branes, the integrals in the area functional are the same as those in asymptotically $AdS_{4}$ and $AdS_{7}$ studied in~\cite{Bhattacharya:2012mi} respectively. This is not surprising as both of them can be obtained by dimensional reduction on the 11-th dimension of M2- and M5-branes, which are asymptotically $AdS_{4}$ and $AdS_{7}$.
\end{itemize}
\subsection{The sphere}
In this subsection we consider the case when the entangling region is a sphere $\sum\limits^{p}_{i=1}x_{i}^{2}\leq l^{2}$, in order to show that the entanglement temperature still exhibits the universality. Unlike the AdS case where the exact solution of the entangling surface is known, here we have to resort to numerics. The nontrivial examples include D2-branes and D4-branes.

\textbf{D2-branes}

The induced metric is given by
\begin{equation}
ds^{2}_{\rm ind}=\frac{1}{(g_{YM}^{2}Nz)^{1/3}}\left[(g(z)+r^{\prime2}(z))\frac{dz^{2}}{z^{2}}+\frac{r^{2}}{z^{2}}d\phi^{2}\right],
\end{equation}
which leads to the area functional
\begin{equation}
A=2\pi g_{YM}^{-2/3}N^{5/3}\int dz\frac{r(z)}{z^{7/3}}\sqrt{g(z)+r^{\prime2}(z)}.
\end{equation}
The corresponding equation of motion for $r(z)$ is
\begin{equation}
-3z\left[1+r^{\prime2}(z)\right]+r(z)\left[-7r^{\prime}(z)-7r^{\prime3}(z)+3zr^{\prime\prime}(z)\right]=0,
\end{equation}
with the boundary conditions
$$r(z_{\ast})=0,~~r^{\prime}(z_{\ast})=\infty,$$
where $z_{\ast}$ denotes the turning point and we have $r(0)=l$\footnote{The infinity in $r^{\prime}(z)$ can be handled by changing the variable in practical calculations. The prescription is the same for D4-branes.}.
Moreover, according to~(\ref{1sthee}), the first order correction to HEE is given by
\begin{eqnarray}
\Delta S_{D2sph}&=&\frac{\pi}{4G_{N}^{(4)}}g_{YM}^{-2/3}N^{5/3}z_{0}^{-10/3}\int dz\frac{zr(z)}{\sqrt{1+r^{\prime2}(z)}}\nonumber\\
&\approx&\frac{0.23676\pi}{4G_{N}^{(4)}}g_{YM}^{-4}U_{0}^{5}l^{3},
\end{eqnarray}
where we have substituted the numerical solution of $r(z)$ to the integral to obtain the above result.
Furthermore, the energy of excited states is
\begin{equation}
\Delta E=\frac{1}{16\pi G_{N}^{(4)}}\frac{7}{2}g_{YM}^{-4}U_{0}^{5}\pi l^{2},
\end{equation}
which results in the entanglement temperature
\begin{equation}
T_{\rm ent}=\frac{\Delta E}{\Delta S_{D2sph}}\approx\frac{1.17639}{l}.
\end{equation}

\textbf{D4-branes}

The calculations for D4-branes can be performed in a parallel way. Given the induced metric
\begin{equation}
ds^{2}_{\rm ind}=\frac{g_{YM}\sqrt{N}}{z^{1/2}}\left[(g(z)+r^{\prime2}(z))\frac{dz^{2}}{z^{2}}+\frac{r^{2}}{z^{2}}d\Omega_{3}^{2}\right],
\end{equation}
we can obtain the area functional
\begin{equation}
A=g_{YM}^{2}N^{3}\Omega_{3}\int dz\frac{r(z)^{3}}{z^{5}}\sqrt{g(z)+r^{\prime2}(z)}.
\end{equation}
Note that unlike the strip case, the integral is not the same as the AdS counterpart any more. The corresponding equation of motion for $r(z)$ is given by
\begin{equation}
-3z\left[1+r^{\prime2}(z)\right]+r(z)\left[-5r^{\prime}(z)-5r^{\prime3}(z)+zr^{\prime\prime}(z)\right]=0,
\end{equation}
with the following boundary conditions
$$r(z_{\ast})=0,~~r^{\prime}(z_{\ast})=\infty$$
and we have $r(0)=l$.
Upon substituting the numerical solution of $r(z)$ into the integral, the first order correction to HEE is given by
\begin{eqnarray}
\Delta S_{D4sph}&=&\frac{\Omega_{3}}{8G_{N}^{(6)}}g_{YM}^{2}N^{3}z_{0}^{-6}\int dz\frac{zr(z)^{3}}{\sqrt{1+r^{\prime2}(z)}}\nonumber\\
&\approx&\frac{0.117943\Omega_{3}}{8G_{N}^{(6)}}g_{YM}^{2}N^{3}z_{0}^{-6}l^{5}.
\end{eqnarray}
The energy of excited states reads
\begin{equation}
\Delta E=\frac{1}{16\pi G_{N}^{(6)}}\frac{5}{8}g_{YM}^{-4}U_{0}^{3}l^{4}\Omega_{3},
\end{equation}
which leads to the entanglement temperature
\begin{equation}
T_{\rm ent}=\frac{\Delta E}{\Delta S_{D4sph}}\approx\frac{0.843389}{l}.
\end{equation}

Once again we see that the entanglement temperature exhibits its universality for spherical entangling regions.
\section{A first-law like relation of holographic entanglement entropy}
In this section we derive a more complete version of the first-law like relation of HEE for non-conformal branes. As observed in~\cite{Guo:2013aca, Allahbakhshi:2013rda, Blanco:2013joa}, when the entangling region is a strip, the variation of the entanglement entropy includes other components of the stress energy tensor other than $T_{tt}$. However, their analysis was carried out in asymptotically AdS spacetimes, in which a well-defined holographic renormalization scheme has been established. The holographic renormalization for more general non-conformal backgrounds is not well understood yet, which may obscure our analysis on the entanglement thermodynamics. Fortunately for non-conformal D-branes, the holographic renormalization has been extensively studied in~\cite{Kanitscheider:2008kd}, which enables us to relate the fluctuations of the metric to the components of stress energy tensor, hence a first-law like relation can be obtained.
\subsection{The dual frame and holographic renormalization}
The holographic renormalization of non-conformal D-branes are performed in the dual frame~\cite{Kanitscheider:2008kd}, which was introduced in~\cite{Boonstra:1998mp}. The dual frame is related to the ten-dimensional string frame by a Weyl transformation,
 \begin{equation}
 ds^{2}_{\rm dual}=(Ne^{\phi})^{c}ds^{2}_{\rm str},~~c=-\frac{2}{7-p}.
 \end{equation}
 The metric in the dual frame is exactly $AdS_{p+2}\times S^{8-p}$. It has been argued in~\cite{Boonstra:1998mp} that the dual frame is the holographic frame in the sense that the radial coordinate is identified with the energy scale in the dual field theory.
The lower dimensional effective action is given by
\begin{eqnarray}
S&=&M\int d^{p+2}x\sqrt{-g}e^{\gamma\phi}\left[R+\beta(\partial\phi)^{2}+C\right],\nonumber\\
\gamma&=&\frac{2(p-3)}{7-p},~~\beta=\frac{4(p-1)(p-4)}{(7-p)^{2}},\nonumber\\
C&=&\frac{1}{2}(9-p)(7-p)\mathcal{R}^{2},~~\mathcal{R}=\frac{2}{5-p},\nonumber\\
M&=&\frac{\Omega_{8-p}}{2\pi^{7}\alpha^{\prime4}}r_{0}^{\frac{(7-p)^{2}}{(5-p)}}\mathcal{R}^{\frac{9-p}{5-p}},~~r_{0}^{7-p}=d_{p}Ng_{YM}^{2}.
\end{eqnarray}
The solution is $AdS_{p+2}$ plus a nontrivial dilaton, which can be expressed in FG coordinates as,
\begin{eqnarray}
ds^{2}&=&\frac{dz^{2}}{z^{2}}+\frac{1}{z^{2}}g_{ij}(x,z)dx^{i}dx^{j},\nonumber\\
\phi(x,z)&=&2\alpha\log z+\kappa(x,z)/\gamma,~~\alpha=-\frac{(p-7)(p-3)}{4(p-5)}.
\end{eqnarray}
The asymptotic expansions of the fluctuations of the metric and the dilaton are given by~\cite{Kanitscheider:2008kd}
\begin{eqnarray}
\kappa(x,z)&=&\kappa_{(0)}(x)+z^{2}\kappa_{(2)}(x)+\cdots+z^{2\sigma}\kappa_{(2\sigma)}(x),\nonumber\\
g_{ij}(x,z)&=&g_{(0)ij}(x)+z^{2}g_{(2)ij}(x)+\cdots+z^{2\sigma}g_{(2\sigma)ij}(x),
\end{eqnarray}
where $\sigma=(p-7)/(p-5)$. In analogy with the standard AdS/CFT, here $g_{(0)}$ sources the stress energy tensor and the scalar field $\phi$ determines the (dimensionful) gauge coupling $g_{d}^{2}=1/\Phi_{(0)}$ by
\begin{eqnarray}
\Phi(x,z)&=&\exp (\chi\phi(x,z))=z^{3-p}(\Phi_{(0)}(x)+z^{2}\Phi_{(2)}(x)+\cdots),\nonumber\\
\Phi_{(0)}(x)&=&\exp\left(-\frac{p-5}{p-3}\kappa_{(0)}(x)\right),~~\chi=-2\frac{p-5}{p-7}.
\end{eqnarray}
The stress energy tensor and Ward identities for non-conformal D-branes are listed in a case-by-case manner in the following. Note that $\langle\mathcal{O}\rangle$ denotes the expectation value of the operator dual to $\Phi$ while $\langle\mathcal{O}_{\phi}\rangle$ denotes the counterpart for $\phi$.

\textbf{D1-branes}

Here we have $\sigma=3/2$ and the Ward identities are given by
\begin{equation}
\langle T_{i}^{i}\rangle-2\Phi_{(0)}\langle\mathcal{O}\rangle=0,~~
\nabla^{i}\langle T_{ij}\rangle+\partial_{j}\Phi_{(0)}\langle\mathcal{O}\rangle=0,
\end{equation}
where
\begin{equation}
\langle\mathcal{O}\rangle=-3e^{3\kappa_{(0)}}M\kappa_{(3)},
~~\langle T_{ij}\rangle=3Me^{\kappa_{(0)}}g_{(3)ij},
\end{equation}
with $\Phi_{(0)}=e^{-2\kappa_{(0)}}$.

\textbf{D2-branes}

Here $\sigma=5/3$ and the Ward identities become
\begin{equation}
\langle T_{i}^{i}\rangle-\Phi_{(0)}\langle\mathcal{O}\rangle=0,~~
\nabla^{i}\langle T_{ij}\rangle+\partial_{j}\Phi_{(0)}\langle\mathcal{O}\rangle=0,
\end{equation}
where
\begin{eqnarray}
& &\langle\mathcal{O_{\phi}}\rangle=-3e^{3\kappa_{(0)}}M\kappa_{(3)},~~
\langle\mathcal{O}_{\phi}\rangle=\chi\Phi_{(0)}\langle\mathcal{O}\rangle,\nonumber\\
& &\langle T_{ij}\rangle=\sigma Me^{\kappa_{(0)}}g_{(2\sigma)ij},
\end{eqnarray}
with $\Phi_{(0)}=e^{-3\kappa_{(0)}}$.

\textbf{D4-branes}

The situation for D4-branes becomes a little complicated as there are logarithmic terms in the FG expansion,
\begin{eqnarray}
g(x,z)&=&g_{(0)}(x)+z^{2}g_{(2)}x+z^{4}g_{(4)}(x)+z^{6}g_{(6)}(x)+2z^{6}\log zh_{(6)}(x)+\cdots,\nonumber\\
\kappa(x,z)&=&\kappa_{(0)}(x)+z^{2}\kappa_{(2)}(x)+z^{4}\kappa_{(4)}(x)+z^{6}\kappa_{(6)}(x)+2z^{6}\log z\tilde{\kappa}_{(6)}(x)+\cdots.
\end{eqnarray}
Here we have $\sigma=3$ and the Ward identities are given by
\begin{equation}
\langle T_{i}^{i}\rangle+\Phi_{(0)}\langle\mathcal{O}\rangle=-2Me^{\kappa_{(0)}}a_{(6)},~~
\nabla^{i}\langle T_{ij}\rangle+\partial_{j}\Phi_{(0)}\langle\mathcal{O}\rangle=0,
\end{equation}
where
\begin{eqnarray}
& &\langle\mathcal{O_{\phi}}\rangle=-Me^{\kappa_{(0)}}(8\varphi+\frac{44}{3}\tilde{\kappa}_{(6)}),~~
\langle\mathcal{O}_{\phi}\rangle=\chi\Phi_{(0)}\langle\mathcal{O}\rangle,\nonumber\\
& &\langle T_{ij}\rangle=Me^{\kappa_{(0)}}(6t_{ij}+11h_{(6)ij}).
\end{eqnarray}
$g_{(6)ij}$ and $\kappa_{(6)}$ are determined by the following expressions
\begin{eqnarray}
g_{(6)ij}&=&A_{ij}-\frac{1}{24}S_{ij}+t_{ij},\nonumber\\
\kappa_{(6)}&=&A-\frac{1}{24}S-2\kappa_{(2)}\kappa_{(4)}-\frac{2}{3}\kappa_{(2)}^{3}+\varphi,
\end{eqnarray}
where $\Phi_{(0)}=e^{\kappa_{(0)}}$ $S_{ij}, S, A_{ij}, A, h_{(6)}, \tilde{\kappa}_{(6)}$ are complicated functions of $g_{(n)ij}$ and $\kappa_{(n)}$, $n=0,2,4$. For details see Section 5.6 of~\cite{Kanitscheider:2008kd}.
\subsection{The first-law like relation}
Given the expressions for the stress energy tensor, it is straightforward to carry out the analysis on the entanglement thermodynamics. As indicated by our previous calculations, it is sufficient to consider only fluctuations of the metric. Therefore the first order correction to HEE is given by
\begin{equation}
\label{dualhee}
\Delta S_{A}=2\pi M\int d^{p}x\sqrt{{\rm det}G_{\alpha\beta}^{(0)}}G_{\alpha\beta}^{(1)}G^{(0)\alpha\beta}e^{\gamma\phi}.
\end{equation}
Note that the factor $e^{\gamma\phi}$ is included to ensure that the evaluation of the minimal surface area is performed in Einstein frame, just as the cases in~\cite{Ryu:2006ef, vanNiekerk:2011yi}. Moreover, we assume that the only nonvanishing component of the fluctuations is $g_{(2\sigma)ij}$. After imposing all these assumptions the expressions for $T_{ij}$ and Ward identities are simplified considerably,
\begin{eqnarray}
\label{ward}
T_{ij}&=&2\sigma Mg_{(2\sigma)ij},\nonumber\\
T_{i}^{i}&=&0,~~\nabla^{i}T_{ij}=0.
\end{eqnarray}
In the following we consider the strip case.

\textbf{D1-branes}

The first order correction to HEE is given by
\begin{eqnarray}
\Delta S_{A}&=&\frac{2}{3}\pi\int dxz\sqrt{1+z^{\prime2}}\left(T_{tt}-T_{xx}\frac{z^{\prime2}}{1+z^{\prime2}}\right)\nonumber\\
&=&\frac{\pi^{2}}{3}z_{\ast}^{2}\left(T_{tt}-\frac{1}{2}T_{xx}\right),
\end{eqnarray}
where we have applied~(\ref{dualhee}), the expressions for $T_{ij}$ and the traceless condition of the stress energy tensor in~(\ref{ward}).
Next recall the expression for $l$ in~(\ref{ld1}), we obtain
\begin{equation}
\label{1std1}
\Delta E=T_{\rm ent}\Delta S_{A}+\frac{1}{2}\Delta PV,
\end{equation}
where
\begin{eqnarray}
& &\Delta E=T_{tt}l,~~\Delta P=T_{xx},\nonumber\\
& &T_{\rm ent}=\frac{12\Gamma(\frac{3}{4})^{2}}{\pi\Gamma(\frac{1}{4})^{2}}\frac{1}{l}.
\end{eqnarray}

\textbf{D2-branes}

The calculations for D2-branes are straightforward,
\begin{equation}
\Delta S_{A}=\frac{3}{5}\pi L\int dxz\sqrt{1+z^{\prime2}}\left(T_{tt}-T_{xx}\frac{z^{\prime2}}{1+z^{\prime2}}\right).
\end{equation}
Next recall the expression for $l$ in~(\ref{ld2}), we obtain
\begin{equation}
\label{1std2}
\Delta E=T_{\rm ent}\Delta S_{A}+\frac{7}{13}\Delta PV,
\end{equation}
where
\begin{eqnarray}
& &\Delta E=T_{tt}lL,~~\Delta P=T_{xx},\nonumber\\
& &T_{\rm ent}=\frac{5\Gamma(\frac{13}{14})\Gamma(\frac{5}{7})^{2}}{14\sqrt{\pi}\Gamma(\frac{3}{7})\Gamma(\frac{17}{14})^{2}}\frac{1}{l}.
\end{eqnarray}

\textbf{D4-branes}

For D4-branes we have
\begin{equation}
\Delta S_{A}=\frac{1}{3}\pi L^{3}\int dxz\sqrt{1+z^{\prime2}}\left(T_{tt}-T_{xx}\frac{z^{\prime2}}{1+z^{\prime2}}\right).
\end{equation}
We can arrive at the following result by using the expression for $l$~(\ref{ld4}),
\begin{equation}
\label{1std4}
\Delta E=T_{\rm ent}\Delta S_{A}+\frac{5}{7}\Delta PV,
\end{equation}
where
\begin{eqnarray}
& &\Delta E=T_{tt}lL^{3},~~P=T_{xx},\nonumber\\
& &T_{\rm ent}=\frac{60\Gamma(\frac{7}{10})\Gamma(\frac{3}{5})^{2}}{\sqrt{\pi}\Gamma(\frac{1}{5})\Gamma(\frac{1}{10})^{2}}\frac{1}{l}.
\end{eqnarray}

\textbf{Remarks}

Here are some remarks on the results:
\begin{itemize}
\item For non-conformal D-branes we still have the first-law like relation for the HEE, given by~(\ref{1std1}),~(\ref{1std2}) and~(\ref{1std4}).
\item The entanglement temperature agrees with previous results up to numerical factors, which is due to different normalizations in the backgrounds.
\item For asymptotically $AdS_{d+1}$ spacetimes the first-law like relation is given by~\cite{Allahbakhshi:2013rda}
\begin{equation}
\Delta E=T_{\rm ent}\Delta S_{A}+\frac{d-1}{d+1}\Delta PV.
\end{equation}
It can be seen that our results for D1- and D4-branes agree with the above expression with $d=3$ and $d=6$ respectively. which can be attributed to their connections to M-branes.
\end{itemize}
\section{Summary and discussion}
Investigations on thermodynamics of entanglement were initiated in~\cite{Bhattacharya:2012mi}, where the authors derived a first-law like relation for the holographic entanglement entropy of excited states. In particular, the entanglement temperature $T_{\rm ent}$ scales as $T_{\rm ent}=c/l$, where $c$ is a dimensionless constant which depends only on the shape of the entangling region. Subsequently, it was observed that when the entangling region is a strip, other components of the stress energy tensor besides $T_{tt}$ also appear in the first-law like relation when translational invariance is broken~\cite{Guo:2013aca, Allahbakhshi:2013rda, Blanco:2013joa}. All the above mentioned analysis was performed in asymptotically AdS spacetimes, which means that the dual description is excited states of CFTs. Therefore it would be interesting to see if the first-law like relation still holds and $c$ is solely dependent on the geometry of the entangling region in non-conformal backgrounds.

In this paper we study entanglement thermodynamics for non-conformal Dp-branes, $p=1,2,4$, which are natural candidates for investigations in non-conformal backgrounds. Our main results are summarized as follows:
\begin{itemize}
\item We calculate the entanglement temperature of a strip using ten-dimensional string frame metric and find that $c$ is also uniquely determined by the shape of the entangling region and is independent of any coupling.
\item To check if such a behavior is universal we work in the $(p+2)$-dimensional effective metric and observe that $c$ is still uniquely determined by the shape of the entangling region.
\item We also consider the case in which the entangling region is a sphere and calculate the entanglement temperature by numerics. It turns out that $c\sim O(1)$.
\item When more general fluctuations of the metric are turned on, the first-law like relation receives contributions from other components of the stress energy tensor, just like the asymptotically AdS cases.
\item The first-law like relations for D1- and D4-branes are exactly the same as the counterparts in asymptotically $AdS_{4}$ and $AdS_{7}$, which may be attributed to their connections to M2- and M5-branes.
\end{itemize}

A closely related subject to the entanglement thermodynamics is the relative entropy defined as
\begin{equation}
S(\rho_{1}|\rho_{0})={\rm tr}(\rho_{1}\log\rho_{1})-{\rm tr}(\rho_{1}\log\rho_{0}),
\end{equation}
where $\rho_{0}$ and $\rho_{1}$ are two density matrices. The relative entropy is always positive and increasing with the size of the system. It was observed in~\cite{Blanco:2013joa} that for two states which are infinitesimally close to each other, vanishing of the relative entropy results in the following equality
\begin{equation}
\label{sh}
\Delta S=\Delta H,
\end{equation}
where $\Delta S$ denotes the first order variation of the entanglement entropy and $\Delta H$ is the first order variation of the expectation value of the \textit{modular} Hamiltonian. It was argued in~\cite{Blanco:2013joa} that the origin of the entanglement temperature is simply~(\ref{sh}). For non-spherical entangling regions where the modular Hamiltonian is not explicitly known, requiring~(\ref{sh}) to hold may suggest the appearance of new operators in the modular Hamiltonian. Therefore our results may provide insightful information for the modular Hamiltonian for non-conformal field theories.

Another interesting subject is to study the dynamics of entanglement entropy and the behavior of entanglement density in these non-conformal D-branes backgrounds, following~\cite{Nozaki:2013vta, Nozaki:2013wia, Bhattacharya:2013bna}. There the authors considered a spherical entangling region in asymptotically AdS, where the minimal surface is explicitly known. Unfortunately, this is not the case any more for non-conformal D-branes. However, due to the connections between D1-, D4- and M2-, M5-branes, it may be reasonable to expect that the dynamics of entanglement entropy of the lower dimensional systems can be inferred from their higher-dimensional asymptotically AdS origins. We hope to report progress on this topic soon.

\bigskip \goodbreak \centerline{\bf Acknowledgments}
\noindent
We thank Tadashi Takayanagi and Marika Taylor for valuable discussions. Part of this work was delivered at the workshop ``Holography: From Gravity to Quantum Matter'' at Isaac Newton Institute for Mathematical Sciences, University of Cambridge. We also thank the participants, especially Hong Liu, Robert Myers, Sang-Jin Sin, Brian Swingle, Erik Tonni and Javier Molina Vilaplana for interesting discussions and comments. DWP is supported by Alexander von Humboldt Foundation.

\newpage

\end{document}